\documentclass[12pt]{article}

\usepackage{bm}
\usepackage[T2A]{fontenc}
\usepackage[english]{babel}
\usepackage{amsfonts,amssymb,amsmath}
\usepackage{amsmath,graphicx}
\usepackage{graphicx,pifont} 
\usepackage{epsfig}

\begin{document}
%
\title{Modeling of hysteresis phenomena in crystalline ferroelectrics: hysteresis loops shape control by means of electric field parameters}
\author{A.Yu.~Zakharov, M.I.~Bichurin\\  \\ \it 
Yaroslav-the-Wise Novgorod State University\\
\it Velikiy Novgorod, 173003, Russia \\ \\
e-mail:  Anatoly.Zakharov@novsu.ru}

\date{}
\maketitle

\begin{abstract}
Description of domains switching processes in crystalline ferroelectrics by means of relaxation type equations with account of variable external electric field is proposed. 
Exact solution of these equations in closed analytical form at arbitrary dependence of electric field on time is obtained. 
A connection between frequency of sinusoidal external field and hysteresis loops shapes is analysed by numerical way. The calculation results are in qualitative accordance with experimental data. 

\end{abstract}
%



%
\section{Introduction}

The relaxation model of domains switching processes in crystalline ferroelectrics~\cite{ZBYP} based on the following assumptions
\begin{enumerate}\label{hypot}
\item The switching of a single domain probability~$\alpha$ from a metastable to stable state depends on external electric field strength~$\mathbf{E}$ and describes by arbitrary given monotone increasing function~$\alpha \left( \mathbf{E} \right)$. 
\item The switchings of the domains occur independently from each other.
\item The connection~$p\left( E\right)$ between single domain order parameter~$p$ and external field strength~$E$ is known (including the metastable states).
\end{enumerate}

In this model, at increasing of external field frequency most of the switchings occur when external fields values belong neighbourhood of metastability boundary. It leads to change of hysteresis curves to the rectangular shape~\cite{Zak4}.

But, part of experimental data~\cite{Priya1,Priya2} are contrary to this conclusion. 

The present paper contains a development of the relaxation model to improve some contradiction between model~\cite{ZBYP} and part of experimental data on hysteresis loops.  
It is well known that variable external field initiates decay of metastable states (such as overcooled liquids avalanche crystallization or supersaturated solutions decomposition). It means that decay rates of metastable states depends on both external field value and its derivative with respect to time. 

This paper devoted to account of both instantaneous values external electric field~$E(t)$  and its derivative with respect to time~$E'(t)= \frac{dE(t)}{dt}$ in relaxation model of switching processes in crystalline ferroelectrics.

\section{The model description}

In addition to basic assumptions of relaxation model of switching processes in crystalline ferroelectrics~\cite{ZBYP} suppose the domains switching probability per unit time~$\alpha\left(E(t), E'(t)\right)$ has the following form
\begin{equation}\label{alpha12}
\alpha\left(E(t), E'(t)\right) = \alpha_1\left(E(t) \right)  + \alpha_2 \left (E'(t) \right).
\end{equation}
Suppose both functions~$\alpha_1\left(E(t) \right) $ and $\alpha_2\left(E'(t) \right) $ satisfy conditions
\begin{enumerate}
\item $\alpha_i(x) = \alpha_i(-x)$;
\item both functions $\alpha_i(x)$ are monotone nondecreasing function of $|x|$;
\item $\alpha_i(0)=0$.
\end{enumerate}

Let's choose one of the possible directions of polarization of dipoles as the positive direction. Then at~$-E_0 <E \left (t \right) <E_0 $ ($E_0$~is the metastability boundary) some fraction of domains belong in the metastable state and pass into the stable states. Then evolution of the domains directions in time obeys the following system of the ordinary differential equations
\begin{equation}\label{evolution}
\left\{
\begin{array}{l}
{\displaystyle  \frac{dn_1(t)}{dt} = \alpha\left( {E(t), E'(t)} \right)\left[ -\theta\left(- {E(t)} \right) n_1(t) + \theta \left( {E(t)} \right) n_2(t) \right];   }\\  
{\displaystyle n_1(t) + n_2(t) = \mathrm{const}  = 1, }
\end{array}
\right.
\end{equation}
where~$n_1\left(t \right)$ and $n_2\left(t \right)$ are fractions of positively and negatively oriented domains respectively, 
 $\theta\left( x\right)$~is the Heaviside step-function given as:
\begin{equation}\label{heavy}
\theta \left( x \right) =   \left\{
\begin{array}{l}
{\displaystyle  1, \qquad x > 0;  }\\
{\displaystyle 0.5,\, \quad x = 0;   }\\
{\displaystyle 0, \qquad x < 0.   }
\end{array}
\right.
\end{equation}
Unlike the equations of evolution~\cite{ZBYP} function~$\alpha\left( {E(t), E'(t)} \right)$ in~(\ref{evolution}) depends not only on instantaneous values of external field~$E(t)$, but also on its derivative~$E'(t)$.

Excluding in the system of equations~(\ref {evolution}) the function $n_2 \left (t \right) $, we obtain the following  closed equation with respect to fraction of positively oriented domains~$n_1\left(t \right)$:
\begin{equation}\label{dn1}
    \frac{dn_1\left( t\right)}{dt} + \alpha\left( {E}(t), E'(t) \right)\, n_1\left( t\right) = \alpha\left( {E}(t), E'(t) \right)\, \theta \left( {E}(t) \right).
\end{equation}
Solution of this equation is simple and has the following form
\begin{equation}\label{n1tdn1t}
\begin{array}{l}
    {\displaystyle      n_1\left( t\right) = e^{-F\left( t\right) } \left[ n_1^{(0)} +  \int\limits_0^t dt_1\ \theta \left ( {E} \left(t_1 \right) \right )\, \alpha \left ( {E}\left(t_1 \right), E'(t_1) \right )  
    e^{  F\left( t_1 \right) } \right],  }
\end{array}
\end{equation}
where~$n_1^{(0)} =  \left. n_1\left(t \right) \right|_{t=0}$ is initial value of function~$n_1(t)$,
and~$F\left( t \right)$ is nondecreasing function defined by relation
\begin{equation}\label{F(t)} 
    F\left( t \right) = \int\limits_{0}^{t} \alpha\left( {E}\left( t_1\right), E'(t_1) \right)\, d t_1.
\end{equation}

The function $n_2\left(t \right)$ can be expressed via $n_1\left(t \right)$ as:
\begin{equation}\label{n2(t)}
    n_2\left(t \right) = 1 - n_1\left(t \right).
\end{equation}

Let's assume that the controling external field~$E \left (t \right) $ is a periodic function with period~$ T $. We shall assume also that on first half of the period~$ \left [0; T/2 \right] $ the function~$ E \left (t \right) $ is positive, and on the second half period~$ \left [T/2; T \right] $ this function is negative. For example, for a sinusoidal controling field
\begin {equation} \label{sin-ot}
{E} \left (t \right) = a \, \sin \left (\omega t \right)
\end {equation}
Function ~ $ {E} \left(t \right) $ is positive in the interval~$\left(0; \pi / \omega  \right) $ and negative in the interval~$\left( \pi / \omega; 2 \pi / \omega \right) $.

In general case of \emph{arbitrary} periodic controling field, we have the solution of the equation~(\ref {dn1}) in the interval~$ \left [0; \ \frac {T} {2} \right] $  
\begin {equation} \label {T1}
n_1 (t) = 1 - \left (1-n_1 ^ {(0)} \right) e ^ {-F (t)}, \quad 0 \le t \le \frac {T} {2}
\end {equation}
and in the interval~$ \left [\frac {T} {2}; \ T \right] $
\begin {equation} \label {T2}
n_1 (t) = \left [e ^ {F \left (\frac {T} {2} \right)} - \left (1-n_1 ^ {(0)} \right) \right] e ^ {-F (t)}, \quad \frac {T} {2} \le t \le T.
\end {equation}
The continuity of solutions~$n_1(t)$ at the point~$\frac {T} {2} $ here is taken into account.

Using these relations we obtain
\begin{equation}\label{n10T}
\begin{array}{rcl}
{\displaystyle  n_1(t)} & = & {\displaystyle \theta\left(\frac{T}{2} - t \right) \left\{  1 - \left( 1-n_1^{(0)} \right) e^{-F(t)} \right\} }\\
{\displaystyle } & + & {\displaystyle  \theta\left(t - \frac{T}{2}\right) \left\lbrace \left[e^{F\left(\frac{T}{2}\right)} -\left( 1-n_1^{(0)}  \right)  \right] e^{-F(t)} \right\rbrace 
  }.
\end{array}
\end{equation}

This solution~$n_1(t)$ is a periodic function  if and only if holds the condition
\begin{equation}\label{period1}
n_1(T) = n_1^{(0)}.
\end{equation}

Thus, within the interval~$ \left [0; T \right] $ the periodic solution of  equation~(\ref {dn1}) has the form~ (\ref {n10T}) with the initial condition~ (\ref {period1}) and function~$ F (t) $, defined by a relation~(\ref{F(t)}).

\section{Dynamics of switching processes in a sinusoidal controling field}

Let's suppose, that functions~$\alpha_i (x) $ in~(\ref {alpha12}) can be approximated by polynomials with coefficients~$ C_k, \ D_k $:
\begin{equation}\label{alpha-i}
\left\{
\begin{array}{l}
{\displaystyle  \alpha_1\left(E(t) \right)  = \sum_{n=1}^{M_1} C_n\, \left(\frac{E(t)}{E_0} \right)^{2n}, } \\
{\displaystyle \alpha_2\left(E'(t) \right)  = \sum_{m=1}^{M_2} D_m\, \left(\frac{E'(t)}{E_0} \right)^{2m},   }
\end{array}
\right.
\end{equation}
where~$E_0$ is the metastability boundary of ferroelectrics.

Substitution~(\ref {sin-ot}) into~(\ref {alpha12}) and~(\ref {F(t)}) leads to integrals which are expressed by elementary functions
\begin{equation}\label{sin2n}
\begin{array}{l}
{\displaystyle  F_1\left(n, t \right) = \int\limits_0^t \sin^{2n}(\omega t)\, dt = \frac{\left( 2n \right)!}{\left[2^n\, n! \right]^2}\, t}\\
{\displaystyle  + \frac{(-1)^n}{2^{2n}\,\omega}\, \sum_{k=0}^{n-1}\, (-1)^k\, C_{2n}^{\, k}\, \frac{\sin\left[2(n-k)\omega t\right]}{n-k},}
\end{array}
\end{equation}
\begin{equation}\label{cos2m}
\begin{array}{l}
{\displaystyle  F_2\left(m, t \right) = \int\limits_0^t \cos^{2m}(\omega t)\, dt = \frac{\left( 2m \right)!}{\left[2^m\, m! \right]^2}\, t}\\
{\displaystyle  + \frac{1}{2^{2m}\,\omega}\, \sum_{k=0}^{m-1}\, C_{2m}^{\, k}\, \frac{\sin\left[2(m-k)\omega t\right]}{m-k}},
\end{array}
\end{equation}
where~$C_{2n}^{\,k} = \frac{(2n)!}{k!\,(2n-k)!}$ are the binomial coefficients.

As a result, the expression~(\ref {F(t)}) for~$F(t)$ consists of linear~$F_L(t)$ and oscillating~$F_O(t)$ terms:
\begin{equation}\label{Lin+Osc}
F(t) = F_L(t) + F_O(t),
\end{equation}
where
\begin{equation}\label{Lin}
F_L(t) = \sum_{n=1}^{M_1} \left[ C_n \left( \frac a 2\right)^{2n} \frac{\left( 2n \right)!}{\left( n! \right)^2}\right]\, t + \sum_{m=1}^{M_2}\left[ D_m \left( \frac{a \omega}{2} \right)^{2m} \frac{\left( 2m \right)!}{\left (  m! \right )^2}\right] \, t ,
\end{equation}
and
\begin{equation}\label{Osc}
\begin{array}{r}
{\displaystyle F_O(t) = \sum_{n=1}^{M_1} \left\lbrace C_n \left( \frac a 2\right)^{2n} \left[ \frac{(-1)^n}{\omega}\, \sum_{k=0}^{n-1}\, (-1)^k\, C_{2n}^{\, k}\, \frac{\sin\left[2(n-k)\omega t\right]}{n-k} \right] \right\rbrace } 
\\ 
{\displaystyle + \sum_{m=1}^{M_2} \left\lbrace D_m \left( \frac{a \omega}{2} \right)^{2m} \left[ \frac{1}{\omega}\, \sum_{k=0}^{m-1}\, C_{2m}^{\, k}\, \frac{\sin\left[2(m-k)\omega t\right]}{m-k} \right]\right\rbrace }.
\end{array}
\end{equation}

Note that
\begin{equation}\label{FO(TT1T2)}
F_O\left( 0\right) = F_O\left( \frac{T}{2}\right) = F_O\left( T\right) = 0.
\end{equation} 
Thus, values of functions~$ F(t)$ and~ $F_L (t) $ at the points~$0,\ \frac T 2,\ T$ coincide, and initial value of function~$ n_1 (t) $ should be fixed by means of relation~(\ref{period1}):
\begin{equation}\label{n10}
\begin{array}{r}
{\displaystyle n_1^{(0)} =  \frac{e^{F\left(\frac{T}{2}\right)}-1}{e^{F\left(T\right)}-1} } \\ \\
{\displaystyle  = {  \left[ 
\exp \left\lbrace {\frac{\pi}{\omega}  \left[ \sum_{n=1}^{M_1} \left( C_n \left( \frac a 2\right)^{2n} \frac{\left( 2n \right)!}{\left( n! \right)^2}\right) + \sum_{m=1}^{M_2}\left( D_m \left( \frac{a \omega}{2} \right)^{2m} \frac{\left( 2m \right)!}{\left (  m! \right )^2}\right) \right] } \right\rbrace  + 1  \right]^{-1} }.}
\end{array}
\end{equation} 
Let's pass to the analysis of possible hysteresis loops shapes within the framework of the relaxation model.

\section{Analysis of hysteresis loops shapes}

Hysteresis curves in variables ``strength of an controlling electric field~$ E (t) $~--- domains directions~$ n_1 (t) $'' are given by parametric representation
\begin{equation}\label{hyst-nE}
\left\{
\begin{array}{l}
{\displaystyle E=E(t);}\\
{\displaystyle n_1 = n_1(t) , }
\end{array}
\right.
\end{equation}
where functions~$E(t)$, $n_1(t)$ are defined by equations~(\ref{sin-ot}) and~(\ref{n10T}) with regard to condition~(\ref{n10}), respectively.
At known values of parameters~$ C_n, \, D_m $, describing ferroelectrics properties, the hysteresis loop shape~$ n_1-E $ depends on both amplitude and frequency of an controling external field. 

Polarization of a ferroelectrics~$P(t)$ at known value~$ n_1 (t) $ can be calculated under the formula
\begin{equation}\label{P(t)}
P(t) = n_1(t)\, p_+(E(t))\ +\ (1-n_1(t))\,p_-(E(t)),
\end{equation}
where~$p_{\pm}(E)$~are two {\em single-valued} functions connecting value dipole moment of a single domain with strength of an external field, including metastable states~\cite{ZBYP}.
The explicit form of these functions depends on the model used for ferroelectrics description.

In the given work the equation of state for ferroelectric with infinite radius of interaction~\cite{ZBE} is used.  This equation coincides with effective field approximation
\begin{equation}\label{E-p}
{E} = \frac{\tau}{2} \ln \left[\frac{1 + p}{1 - p} \right] - p,
\end{equation}
where~$\tau,\ p,\ {E}$~ are dimensionless temperature, order parameter, and external electric field strength, respectively, functions~$p_{\pm}(E)$~are the branches of this equation with respect to~$p$ solution.

\begin{figure}
\includegraphics[width=6.0in]{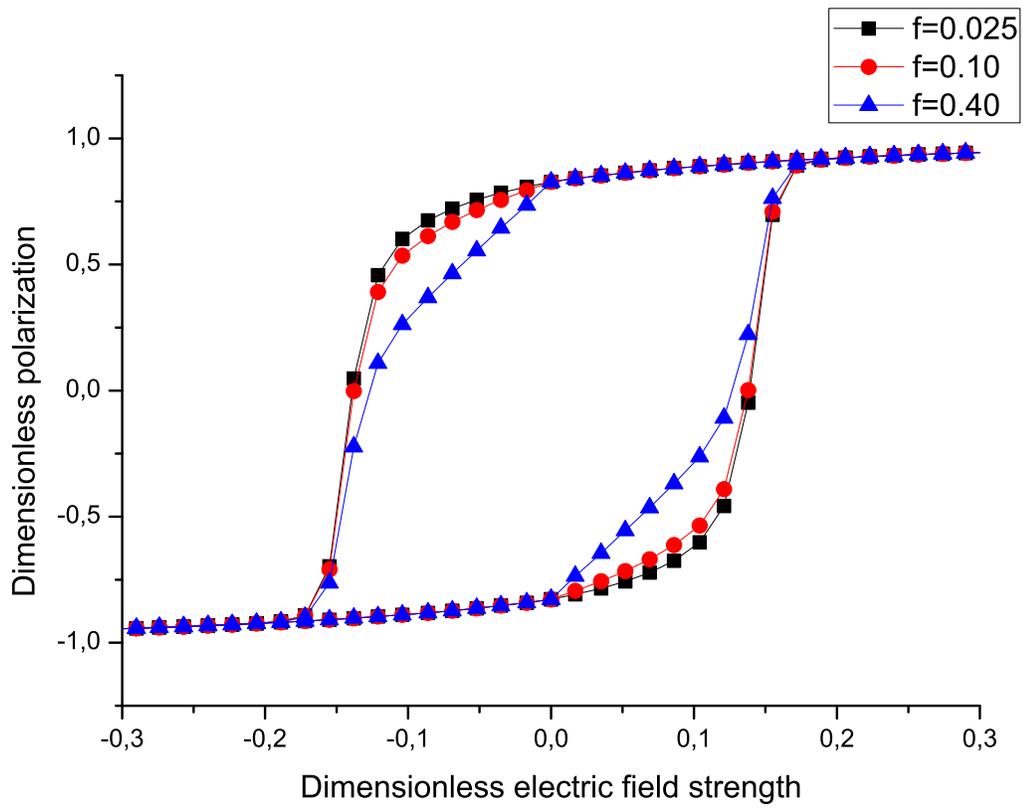}
\caption{Qualitative view of dependence of ferroelectrics hysteresis curves on frequency of an controling external field at $C_1=4.5$, $D_1=1.0$.}.
\label{fig:1}
\end{figure}

\begin{figure}
\includegraphics[width=6.0in]{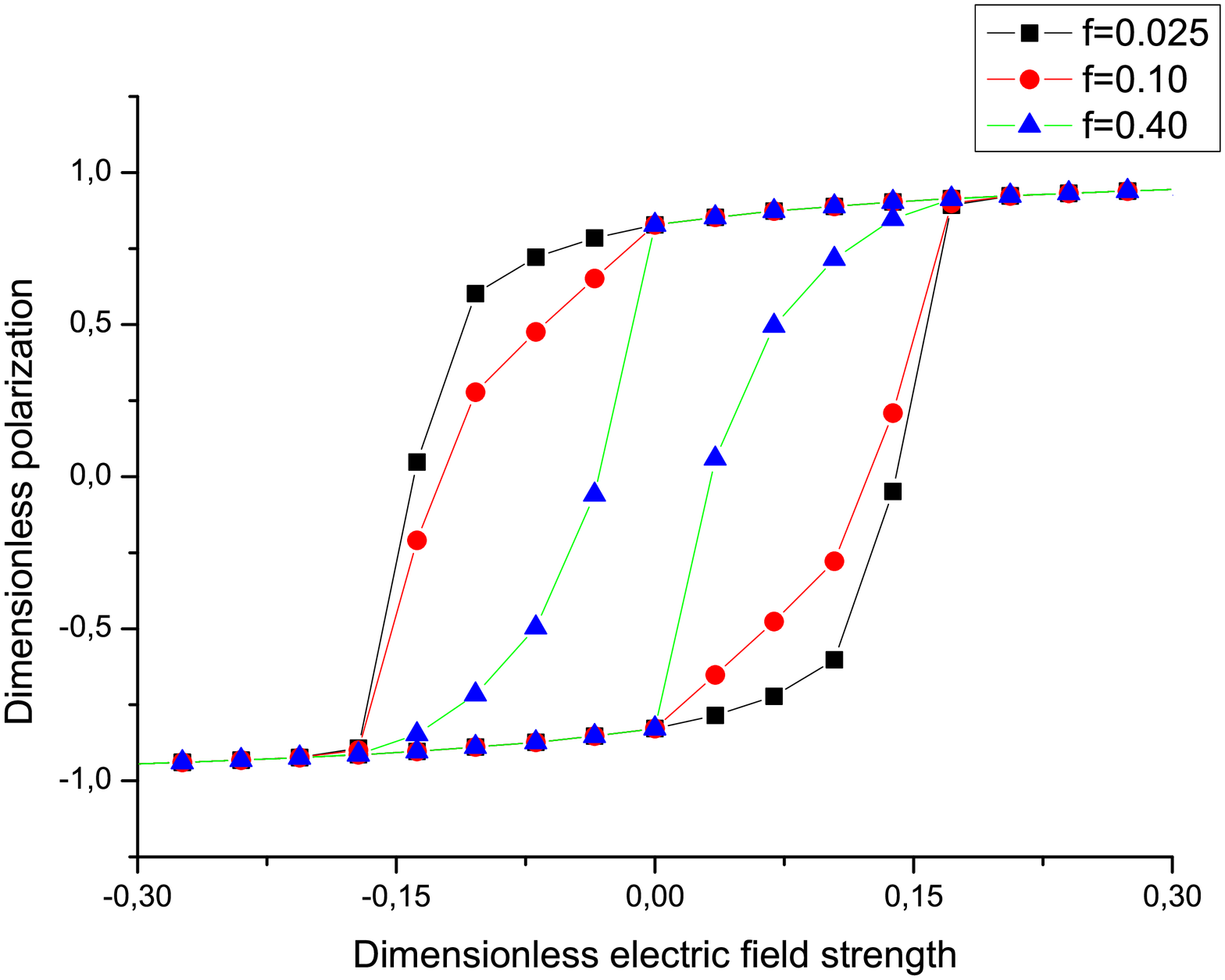}
\caption{Qualitative view of dependence of ferroelectrics hysteresis curves on frequency of an controling external field at $C_2=4.5$ и $D_2=1.0$.}.
\label{fig:2}
\end{figure}

We consider the cases with minimal number of parameters in relations~(\ref {alpha-i}). It is important to trace influence of relation between contributions~$ \alpha_1(E(t)) $ and $\alpha_2 (E ' (t)) $ in~(\ref {alpha12}) on the hysteresis loops shapes at various frequencies of an external field. We shall choose two simplest cases, when each of right hand sides of~(\ref {alpha-i}) contains only one nonzero term:
\begin{enumerate}
\item $C_1$ и $D_1$ (in this case the probability of the domains switching is quadratic function of an external field strength~$\left( E(t)\right)^2 $ and its derivative with respect to time~$\left( E'(t)\right)^2$). Results of numerical calculations at $C_1=4.5$ and $D_1=1.0$ are presented in Fig.\ref{fig:1}.
\item $C_2$ и $D_2$ (in this case the probability of the domains switching contains the terms with fourth degree of an external field strength~~$\left( E(t)\right)^4$ and its derivative~~$\left( E'(t)\right)^4$). Results of numerical calculations at $C_2=4.5$ and $D_2=1.0$ are presented in Fig.\ref{fig:2}.

\end{enumerate}

\section{Conclusion}

The basic results of the paper consist in the following.

\begin{enumerate}
\item The phenomenological model of domain switching processes in crystalline ferroelectrics with account dependence of switching probability on both  instantaneous values of external field~$E(t) $ and on its derivative with respect to time ~$E'(t) $ is proposed.
\item Without taking into account the contribution~$E'(t)$ increase of external field frequency {\em always} transforms the hysteresis loops to the rectangular form. The account of this therm at suitable values of parameters leads to a possibility controls of hysteresis loops form in any direction.
\item It is shown, that the probability of domains switching within the framework of relaxation model should include contributions both an external field~$E(t)$ and its derivative with respect to time~$E'(t)$, as far as {\em both type} dependences of hysteresis loops form on frequency observed in experiments. 
\item The variation of model parameters leads to rather essential modification of hysteresis loops forms on metastable regions of curves and almost does not change the form on stable regions of hysteresis loops.
\item Results of numerical calculations are in the qualitative accordance with known experimental data.
\end{enumerate}

Work is fulfilled by partial financial support Russian Ministry of Education and Science within the framework of a base part.

\newpage

\end{document}